# The structure of the Maxwell spot centroid

**Albert Le Floch** [a,b] , **Guy Ropars** [a,c*]

[a] *Laser Physics Laboratory, University of Rennes, 35042 Rennes cedex, France*
[b] *Quantum Electronics and Chiralities Laboratory, 20 Square Marcel Bouget, 35700 Rennes, France*
[c] *UFR SPM, University of Rennes, 35042 Rennes cedex, France*

**E-Mail address**: guy.ropars@univ-rennes.fr ; albert.lefloch@laposte.net
[*] Corresponding author at: Laser Physics Laboratory, University of Rennes, 35042 Rennes cedex, France.

## Abstract

The dark entoptic Maxwell spot centroids seen through a blue filter, which coincide with the blue cone-free areas centered on the foveas, are shown to exhibit a structure. When observing through the green part of a blue-green exchange filter in a foveascope, after a fixation through the blue part, a small orange disc is seen around the centre of the pale green memory afterimage corresponding to the blue cone-free area. Using artificial pupils with different diameters, we show that this small circular pattern corresponds to the Airy disc due to the Fraunhofer diffraction through the pupil. Typically, for an eye with a 3 mm diameter pupil, the Airy disc exhibits a diameter of about 8 µm at the centre of the usual 100-150 µm Maxwell centroid. Fixation tests show that the towering central maximum of the Airy pattern irradiance corresponds to the preferred locus of the fixation of the eye near the centre of the blue cone-free area of the fovea. The locus of fixation in human vision thus appears to be located in the only area of the fovea where the large chromatic dispersion is cancelled, optimising the eye acuity.

**Keywords**: Entoptic images, Fovea centralis, Maxwell spot, S cones, Chromatic dispersion, Airy pattern, Line of sight

# 1. Introduction

Since the first observation by Maxwell (Maxwell, 1856) of a dark pattern in the blue part of the visual spectrum, the so-called Maxwell's spot has been shown to be composed of three zones (Walls, 1952; Isobe, 1955; Magnussen, 2004). While the whole patterns extend to about 3° in the fovea, the central part which extends to 0.4 to 0.5° have received much attention (Wald, 1967; Williams, 1981). Although it is possible that pigments play a role in the outer part of the Maxwell spot, the absorption of the blue light by pigments is not necessary for the central part, i.e. the Maxwell centroid, since there is no blue photoreceptor in this small area. Indeed, the existence of this centroid is related to the different cone distributions (Marc, 1977), specifically the absence of blue cones in an area sustending about 20-30 arcminutes corresponding to a diameter of 100-150 µm. The existence of this blue cone-free zone was first deduced from psychophysical observations (Wald, 1967; Williams, 1981), and later directly proven post-mortem by staining the blue cones by specific anti-blue opsins (Curcio, 1991; Bumsted, 1999; Martin, 1999; Cornisch, 2004). This anomaly in the cone mosaic of the foveas has intrigued many authors (Walls, 1952; Wald, 1967; Williams, 1981; Calkins, 2001; Le Floch 2010). In fact, the acuity for a normal eye without any refraction problem is limited by diffraction, chromatic dispersion and ultimately by the size and separation of the photoreceptors in the fovea (Smith, 1997). As noted by Wald (Wald, 1967) and Rodiek (Rodiek, 1998), the blue cone-free area cancels the large 1.3 diopter chromatic blue aberration of the eyes. Moreover, as a consequence, one may wonder if this small blue cone-free area could help define the line of sight of the eye and the locus of fixation (Putnam, 2005; Wilk, 2017; Zeffren, 1990; Reiniger, 2021) with its converging eye movements (Rucci, 2015; Intoy, 2020). Indeed, high resolution retinal imaging (Putnam, 2005; Wilk, 2017; Reiniger, 2021) has shown that the locus of fixation was displaced from the location of highest foveal cone density by an average of about 10 arcmin (about 50 µm). The direction of displacements does not appear to be systematic. The deviation from the centre of the free avascular zone also appears random (Zeffren, 1990). Although exploring the entoptic Maxwell centroid is delicate, one may ask if it does not show a finer structure. We can then investigate the potential role of such structure.

# 2. Methods

## *2.1 Subjects*

Different measurements were made for three subjects with normal ocular status. Their ocular dominances were determined using the noise-activated afterimage method (Le Floch, 2017) and also the sighting Miles test. Two have a right-eye dominance and one a left-eye dominance. As the outline of the Maxwell centroid is quasi-circular and more regular for the dominant eye (Le Floch, 2017; see also the Supplementary for the three observers), this study is performed using the dominant eye. The study was performed in accordance with the Declaration of Helsinki.

## *2.2 Method and apparatus*

To introduce the method of observing the structure of the Maxwell centroids, let us specify the area to be explored. When fixating a target, one can define the line of sight of the eye (Fig. 1a) which connects the fixating point to the centre of the entrance pupil and to the fovea (Artal, 2014). The cone distributions on the fovea schematized in Fig 1a (Polyak, 1941; Rodieck, 1998) shows about the centre a small blue cone-free area of about 100 to 150 microns diameter which corresponds to the centroid of the Maxwell spot, appearing as a dark spot when observed through a blue filter (Figs 1b,c). The direct existence of the blue cone-free area has been proven by post-mortem observations of the retinas, after staining the blue cones with specific antibodies (Curcio, 1991; Bumsted, 1999; Martin, 1999; Cornish, 2004).To record the outlines of the Maxwell spot centroids of each observer, we built the first foveascope (Le Floch, 2017). The observer looks successively at a bright white screen using a projector with a brightness of 3000 lumens, through a blue-green exchange filter which optimizes the contrast for the dark centroid observed through the blue part of the filter, by avoiding the 2-3 s fading time of this entoptic image. The transmittance of the filters, with 40 nm bandwidths, centered at 440 nm and 535 nm in the blue and the green respectively, is chosen at 16% for the blue filter and 4% for the green filter. Each observer adjusts the frequency of the movement of the blue-green filter at his convenience between 0.1 Hz to 1 Hz. When using the dominant eye, each observer sees a typical whole Maxwell spot similar to that schematized in Fig. 1b. At first glance, the quasi-circular dark centroid appears as a homogeneous dark area for the three observers, with a mean diameter of about 27 ± 3' for the observer #1 for instance. The two outlines of the Maxwell centroids for each observer are reported in the Supplementary material (Fig. S1).

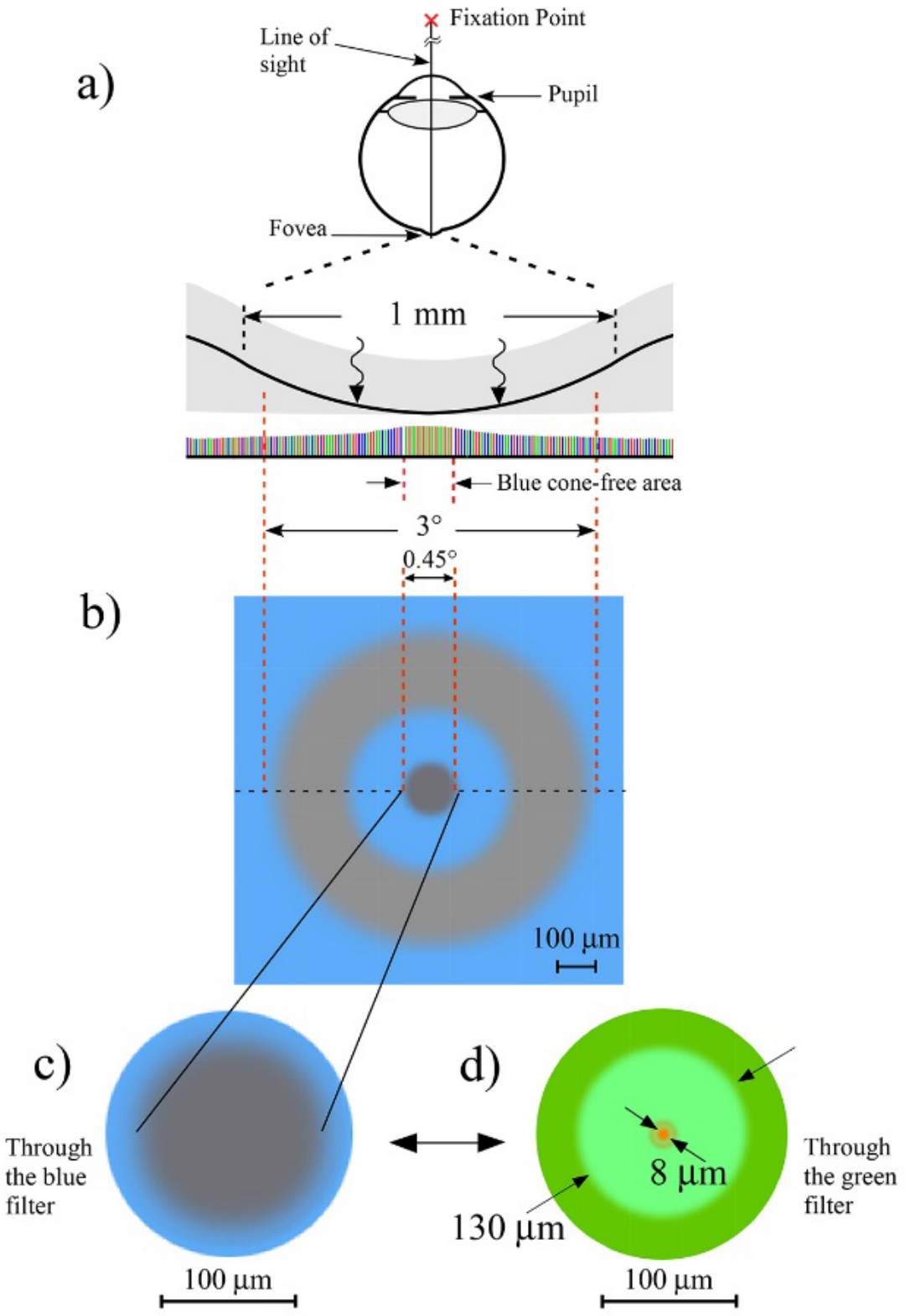

***Figure 1:*** *a) Scheme of an eye with its line of sight and the schematic section through the centre of the human fovea where the photoreceptors are all elongated cones (after Polyak, 1941). b) Scheme of a typical whole Maxwell spot pattern observed through a blue filter (440 nm) for the observer #1. The dark central 0.45° disc, i.e. the Maxwell centroid, corresponds to the blue cone-free area. The dominant eye of the observers has a quasi-circular Maxwell centroid outline (Le Floch, 2017). A similar pattern is seen by all three observers. c) Enlargement of the Maxwell centroid which appears quasi homogeneous. d) When observed through the green filter (540 nm), a circular orange disc is clearly seen about the centre with a faint outer ring (here for the observer #1). Similar discs are seen by all three observers.*

To search for a finer structure of the centroids, we have to focus our attention on the associated light green memory afterimage pattern which can be seen through the green part of the filter, only after fixating through the blue filter (see below Fig. 1d). To investigate more precisely the structure of this green memory afterimage, we have to add some modifications to the foveascope. The experimental set-up remains quite similar to the previous one. The illuminance of the white screen is increased from 5000 to 9000 lux and we use doped glasses to build the blue-green filter. The transmittance of the filters, with 80 nm and 56 nm bandwidths centered at 443 nm and 540 nm in the blue and green respectively, is chosen at 50 % for the blue filter and 17 % for the green filter. To be able to measure the presumed fine structure of the centroid, we project on the screen real small calibrated circles with visual angles from 2 to 5 arcminutes corresponding to 10 to 25 µm diameters on the retina. Moreover, small artificial pupils with diameters from 1 to 3 mm can be interposed just near the pupil of the observer when looking through the foveascope, to investigate the variations of the possible structure. The dark entoptic image intrinsically linked to the blue cone-free area, seen through the blue filter, follows the eye and the artificial pupil movements. By contrast, the light green afterimage seen through the green filter, only observable after a fixation through the blue filter, and corresponding to a memory afterimage encoded in the upper layers of the brain, remains still. Furthermore, to later identify the preferred locus of fixation of the eye we add a 2 mm diameter green LED on the screen which can be lighted at different time intervals.

## 3. Results and discussion

### *3.1 Existence of the structure of the Maxwell centroïd*

When the rather homogeneous dark centroid image shown in Fig. 1c observed through the blue filter is examined through the green filter, the observers see a typical pattern as shown in Fig 1d. For the observer #1 for instance, the 27 $\pm$ 3' diameter dark spot appears as a light green colour afterimage, but now, about its centre, an additional orange disc appears. Besides the perfect circularity of this disc, a faint ring can be observed around the main disc, suggesting a diffraction pattern. Optimizing the frequency of the movement of the blue-green exchange filter around 0.2 Hz facilitates the observation. All the three observers saw a similar structure. Here the Fraunhofer diffraction can be seen, as we are observing another entoptic pattern that requires no more real object, superposed on a memory afterimage. While the dark entoptic Maxwell centroid itself corresponds to the observation of an equivalent real full moon with its angular aperture of 30 arcminutes, the diffraction pattern for a fixation point corresponds to an equivalent real point source of about 1 arcminute of angular aperture. Indeed, in the dynamics of natural vision, the onset of visual fixation affects ongoing neural activity even in the absence of visual stimulation (Rajkai, 2008). Moreover, ocular fixation is known as a dynamic process that is actively controlled (Krauzlis, 2017).

### *3.2 The Airy pattern*

Diffraction is fundamental and unavoidable in most of optical systems such as lasers (Siegman, 1986; Le Floch, 1980; Ropars, 2005), and the eye (Smith, 1997).

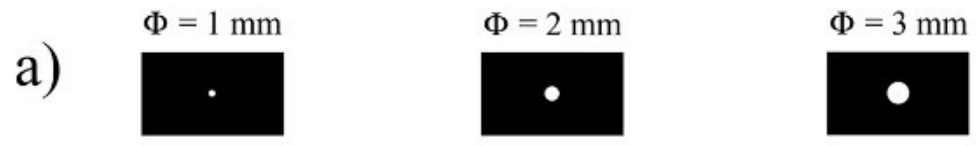

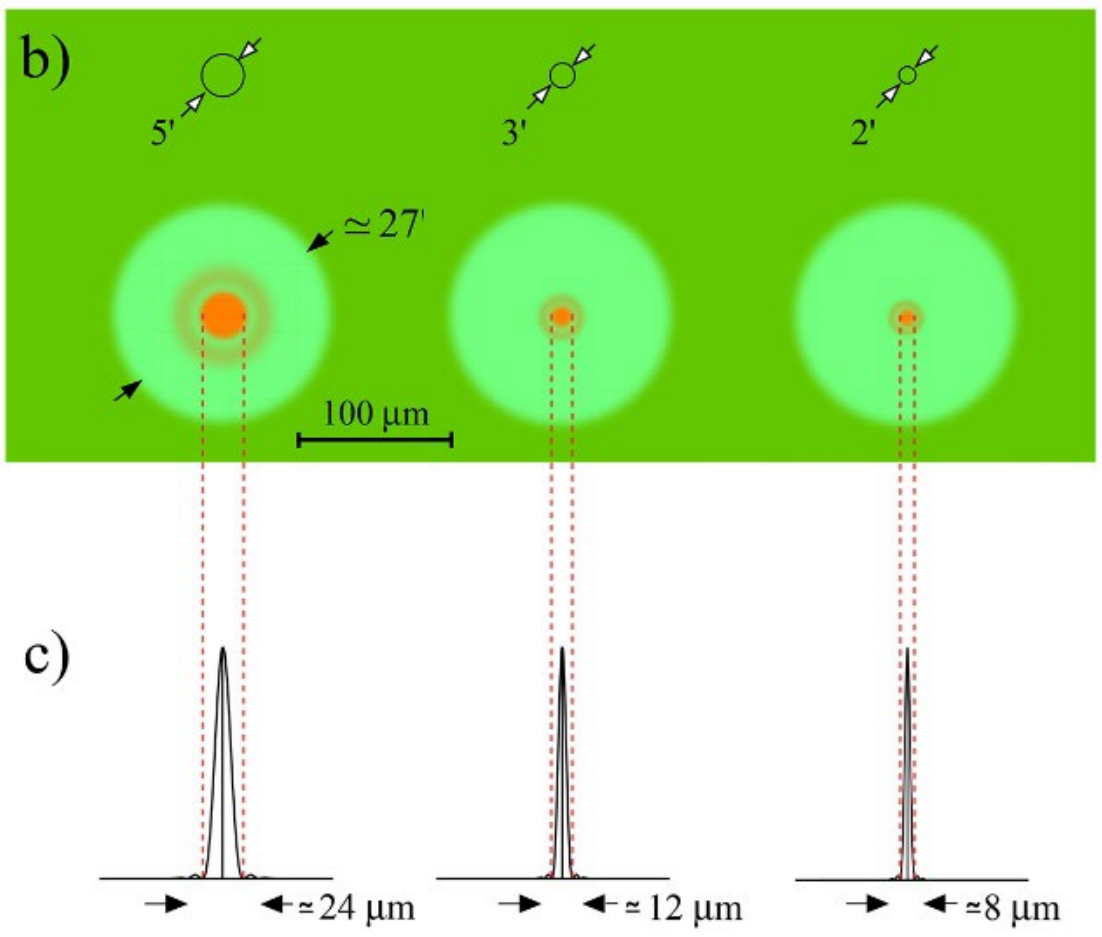

***Figure 2***: *Variations of the size of the orange disc for different artificial pupils. a) The artificial pupils used in front of the eye. b) Observations on the screen of the foveascope through the green filter and comparison with real auxiliary circles (top), for the observer #1. All three observers see similar pattern. c) Theoretical Airy irradiance patterns corresponding to the three artificial apertures for $\lambda$= 540 nm.*

To confirm the role of diffraction, we use three artificial apertures (Fig. 2a) in front of the eye to follow the variations of the orange disk diameter when observing through the foveascope. By comparing the orange disc size to the calibrated real circles projected on the screen of the foveascope (Fig. 2b), the observer can estimate the diameter of the circular disc with a precision of about 10 %. This fine structure of the Maxwell centroid is directly linked to the eye pupil diameter. Indeed, the diameter of the Airy disc is given by the formula (Born, 1999):

$$\Phi = 2.44\,\frac{\lambda f}{d}, \qquad (1)$$

where $f$ is the focal length of the eye ($\simeq 17\,mm$), $d$ the diameter of the pupil, $\lambda \simeq$ 540 nm in the green part of the spectrum. The theoretical transverse distribution of intensity of the light disc in the focal plane for apertures of 1, 2 and 3 mm represented in Fig. 2c are in agreement with the estimated experimental values. The diameter of the orange disc decreases with increasing aperture diameter, following equation (1). The highly contrasted intensity variations of light represented in Fig. 2c, associated with the Airy disc through the green filter with its large bandwidth (56 nm) impinging on the blue cone-free area, can explain the orange colour of the Airy disc. Indeed the bleaching of the green cones is more intense around $\lambda = 540$ nm, corresponding to the maximum transmittance of the filter, than the bleaching of the red cones. This weaker bleaching of the red cones leads to an orange coloured disc observed with the correct filters. Moreover, if a small artificial pupil is gently oscillated in front of the eye, the entoptic orange Airy disc moves synchronously with the aperture across the light green memory afterimage which remains motionless. Note that observing a possible Airy disc directly through the blue filter at the centre of the dark entoptic Maxwell centroid itself is clearly more delicate. However occasionally for some observers a tiny pip of the colour of the background can be discerned in one eye through the blue filter, as previously noted by Wright (Wright, 1953).

### 3.3 Locus of fixation of the eye in the cone topography

The entoptic images are intrinsically linked to a part of the eye. As the entoptic Airy disk moves when oscillating the artificial aperture, and as the other entoptic dark centroid seen through the blue filter also moves synchronously with the aperture, this suggests that the locus of fixation is linked to the blue cone-free area. Let us perform the fixation test using a 2 mm diameter green LED (corresponding to about 2 arcminutes of angular aperture) as a target, as schematized in Fig. 3.

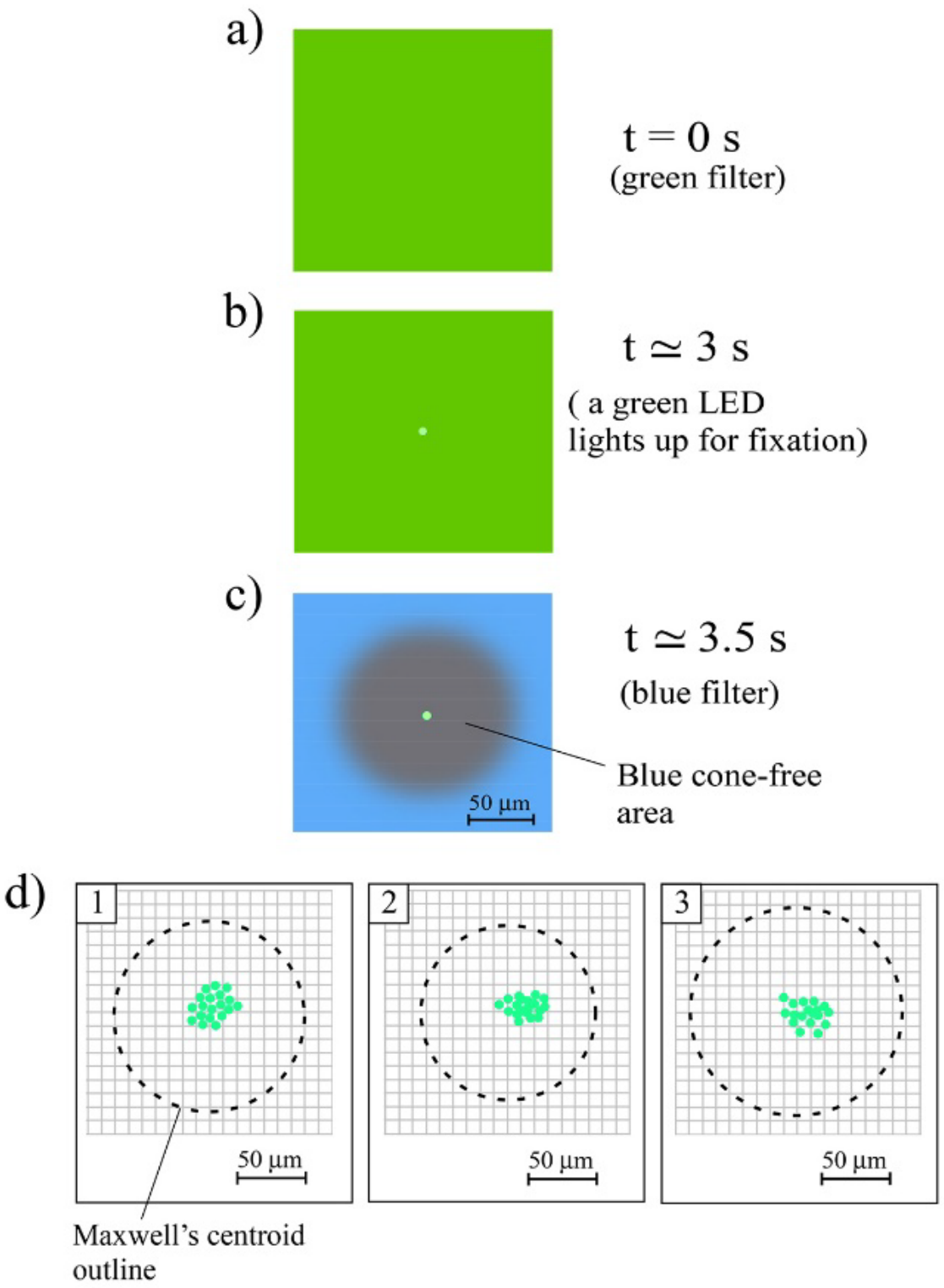

***Figure 3****: Recordings of the monocular preferred fixation locus in the Maxwell centroid profiles using the foveascope. a) Observation of the bright screen through the green filter. b) At t ≃ 3 s, a green LED lights up on the screen for the fixation. c) At t ≃ 3.5 s, the green filter is replaced by the blue filter so as to see the dark entoptic Maxwell centroid, i.e. the blue cone-free area, and localize the fixation LED on the fovea. Here the fixation point is close to the centre, but it can be slightly offset from the centre due to the microsaccades. d) Results for the 20 successive fixation tests for the three observers. The distributions of the fixation points are located about the centre of the Maxwell centroid.*

When the LED lights up (Figs 3a,b), the rapid shift of the exchange filter to the blue part shows that the LED spot is indeed located near the centre of the blue cone-free area, i.e. the dark area representing the Maxwell centroid (Fig. 3c). After 20 successive fixations, the scattering of the fixating points around the centre of the Maxwell centroid shows small deviations (Fig. 3d) for the three observers, with an average of about 6 arcminutes (corresponding to 30 µm), which is in agreement with previous work (Barlow, 1952; Ditchburn, 1953; Putnam, 2005; Wilk, 2017; Reiniger, 2021). However our experiment shows that the stable preferred retinal locus of fixation is linked to the Airy disc about the centre of the blue cone-free area. Indeed, if we shift the exchange filter in the green part after the step schematized in Fig. 3c, the Airy disc appears superposed on the preferred locus of fixation. The scatter of fixation points, also observed by other authors (Putman, 2005, Zeffren, 1990, Wilk, 2017; Reiniger, 2021), but here about the centre of the blue cone free-area, results from the microsaccades (Rucci, 2015 ; Krauzlis, 2017; Intoy, 2020), and from the small movements of the pupil relative to the iris and the eyeball (Nyström, 2013; Mabed, 2014; Bouzat, 2018).

Following the procedure in Fig. 3, we have verified that, for a cohort of 12 students from our University, the preferred retinal locus of fixation is systematically observed about the centre of the blue cone-free area.

### *3.4 Discussion*

Among the main limitations for the eye acuity recalled in the introduction (Smith, 1997) the large 1.3 diopter blue chromatic aberration is the only one which can be cancelled, using the Maxwell centroid i.e. the blue cone-free area. Moreover, the structure observed within this small area represents the Fraunhofer diffraction limitation, but suggests a position for the preferred locus of fixation. It is known that the preferred retinal location is offset from the location of the peak cone density of the fovea by about 50 µm (Putnam, 2005; Wilk, 2017; Reiniger, 2021). It is also offset from the centre of the avascular zone (Zeffren, 1990) and from the pit of the fovea, with no consistency in the direction of the offset across the subjects tested. No clear spatial relationship exists among these foveal specializations and the preferred foveal location (Wilk, 2017). As the blue cone population patterns with the blue cone free-areas are established as soon as 20 weeks after gestation, and as the foveal depression itself only begins to form 5 weeks later (Cornish, 2004), the previous observations of the preferred locus of fixation compared to the other specific regions of the fovea remain compatible with the present locus of fixation about the centre of the blue cone-free area.

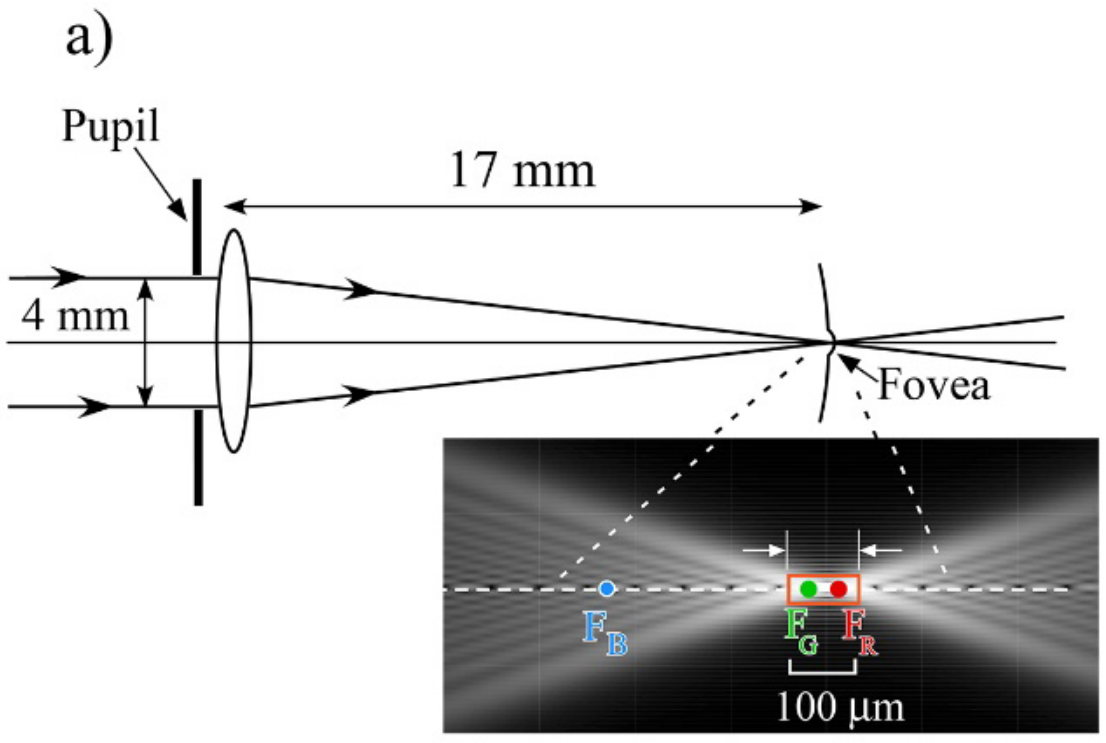

b)

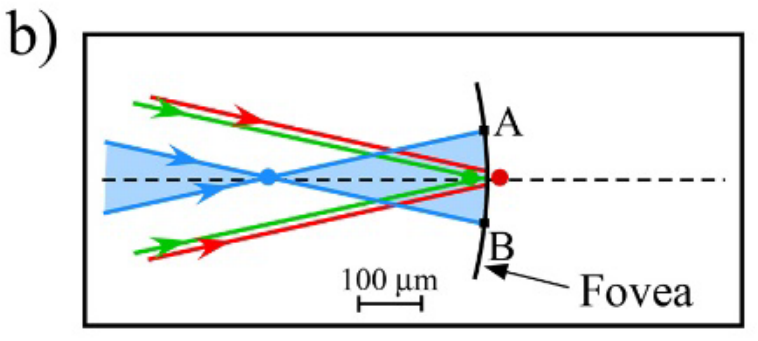

***Figure 4:*** *a) Scheme of a human eye looking at a white screen. The inset represents the focus points on the fovea corresponding to the red and green part of the spectrum. b) Focusing rays pointing to the three types of cones of the fovea. When the eye accommodates for the yellow part of the spectrum (Le Grand, 1967), between the red and green parts of the spectrum, the blurring for the blue rays schematized by AB reaches a 100 μm diameter circle for a pupil with a 4 mm diameter.*

Moreover, one may wonder what are the mechanisms including the eye movements during fixation, which guide the eye fixation toward the centre of the blue cone-free area, where paradoxically the blue part of the visual spectrum cannot be detected. The blue cones with their anatomically distinct pathway for conveying their signals to the brain (Dacey, 1994) play an intriguing role, specially in the fovea with their unique distribution and specificity (Calkins, 2001; Martin, 2014). Their peak spectral sensitivity is well separated (by about 100 nm) from those of the green and red cones, which implies a relatively large defocusing (Rodieck, 1998). The blue image is out of focus for the blue cones as a result of chromatic dispersion, when the eye accommodates about the yellow part of the spectrum (Le Grand, 1967). The typical scheme for the focusing points in the eye for the blue, green and red lights is shown in Fig. 4a. The focus points $F_G$ and $F_R$ for the green and the red part of the visual spectrum respectively are close within the depth of focus of about 100 µm where the spot size of the Airy disc remains the same. In contrast, the focus point for the blue light is located 350 µm before the fovea leading to a bright divergent beam of blue light with a diameter of about 100 µm in the fovea plane (Fig. 4b).

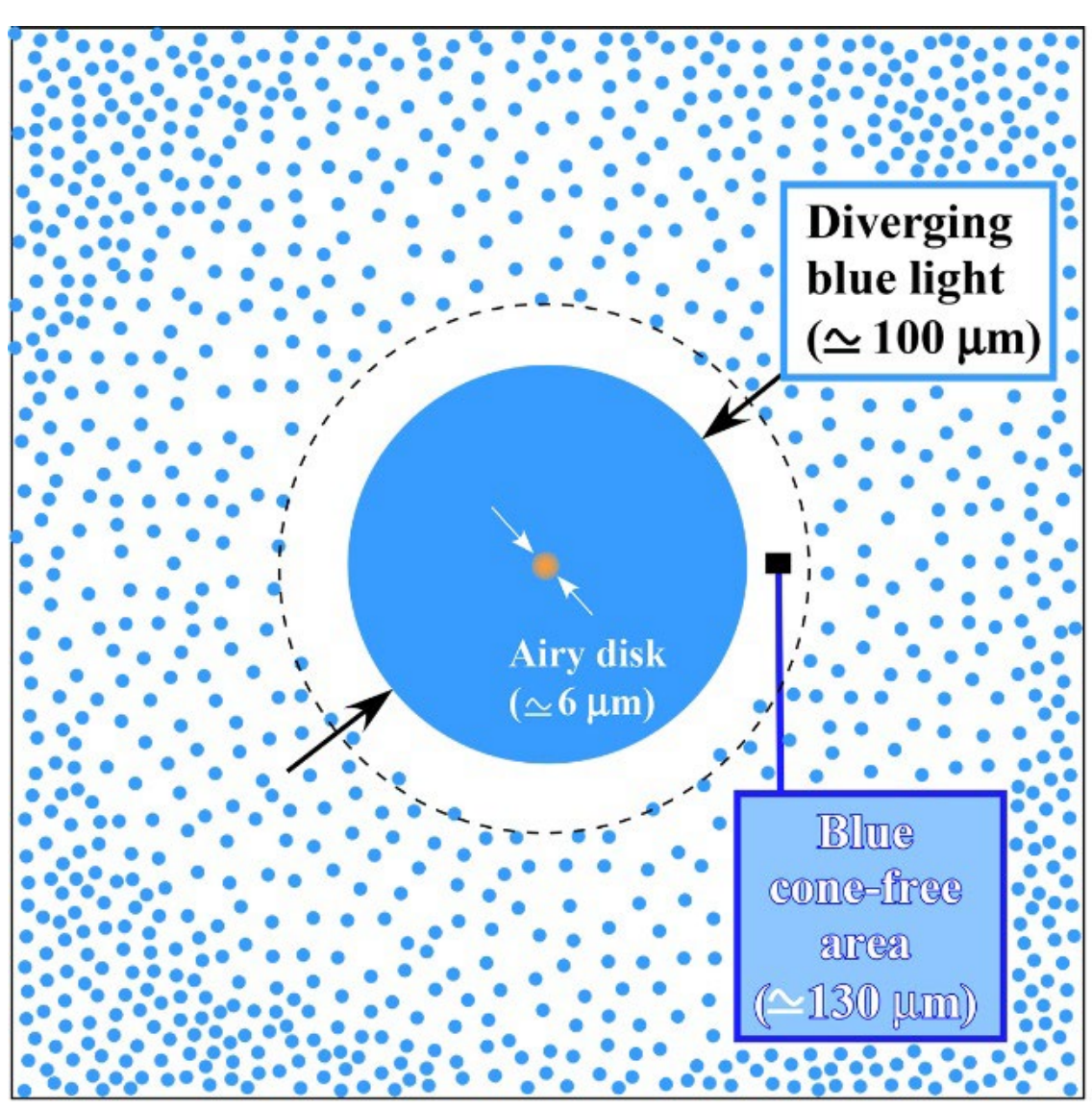

***Figure 5:*** *Scheme of the fovea plane with its blue cone-free area when white light passes through a 4 mm diameter pupil for a stable fixation. The small orange Airy disc around the centre of the blue cone-free area results from the diffraction of the green and red parts of the spectrum, while the blue 100 µm diameter blue zone corresponds to the undetected diverging blue light falling on the blue cone-free area of the fovea.*

When fixing a target the eye seems to be guided toward the fixation locus at the centre of Maxwell centroid, following the Airy disc produced by the circular pupil. However, even during fixation, microsaccades, tremor and shifts occur (Ratliff, 1950; Martinez-Conde, 2002; Mergenthaler, 2007; Rucci, 2015). In the last few years, beyond their role in avoiding image fading (Ditchburn, 1953; Wald, 1967) it has been shown (Rucci, 2015; Intoy and Rucci, 2020) that these small movements essentially contribute to help centering gaze via the superior colliculus and providing a finely controlled oculomotor strategy, reaching 1 arcminute in resolution. As the express saccades and superior colliculus have been shown to be sensitive to the blue short-wavelength cone contrast (Hall and Colby, 2016; Basso, 2016) one may ask if the small contrasted diverging beam of blue light (Fig. 4b) plays a role in the convergence of the eye movements toward the centre of the Maxwell centroid. Indeed when the eye is not correctly fixed at the centre of the centroid, the blue diverging beam falling on the fovea with a 100 µm diameter reaches the first blue cones outside the blue cone-free area (Fig. 5) activating the superior colliculus neurons (Hall and Colby, 2016), suggesting a possible active control of the eye fixational movements converging toward the centre of the blue cone-free area.

### *3.5 Conclusion*

To summarize, the entoptic images and their corresponding memory afterimages appear to bring new insights in vision. The blue cone-free area, an apparent anomaly in the cone mosaic of the fovea seems to be an anatomical feature playing an important role in human vision. It cancels the contribution of the blue radiation at the center of the fovea, thus avoiding the chromatic aberration introduced by Newton, that would cause the greatest problem for acuity and for the image resolution at the locus of fixation. Further, the finer structure of its corresponding Maxwell centroids, linked to the Fraunhofer diffraction Airy disc due to the pupil-lens system on the fovea, determines the locus of fixation and the line sight of the eye. Moreover, the defocused blue diverging beam on the fovea could then be a possible active control of the eye fixational movements converging toward the centre of the blue cone-free area.

**Declaration of Competing Interest**

The authors declare that they have no known competing financial interests or personal relationships that could have appeared to influence the work reported in this paper.

**Acknowledgements**

We like to thank Jean René Thébault for his technical assistance, the observers for their kind participations, Roger Le Naour and Kevin Dunseath for reading the manuscript and for their comments.

**Supplementary**

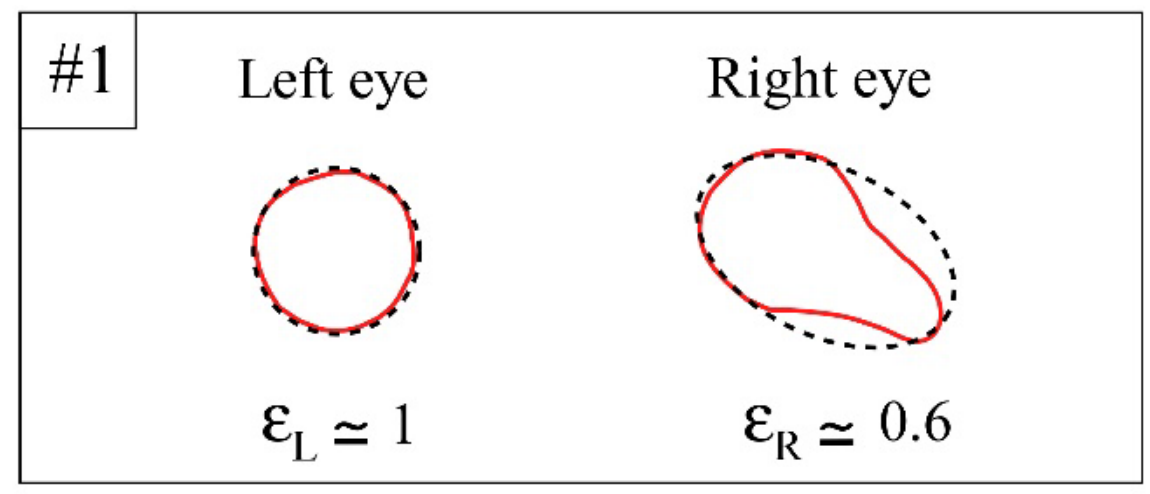

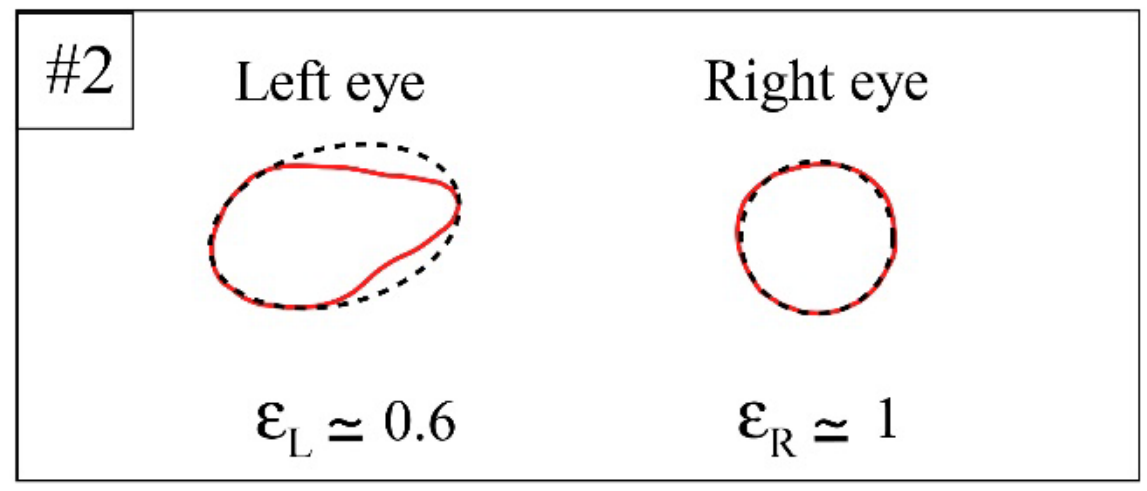

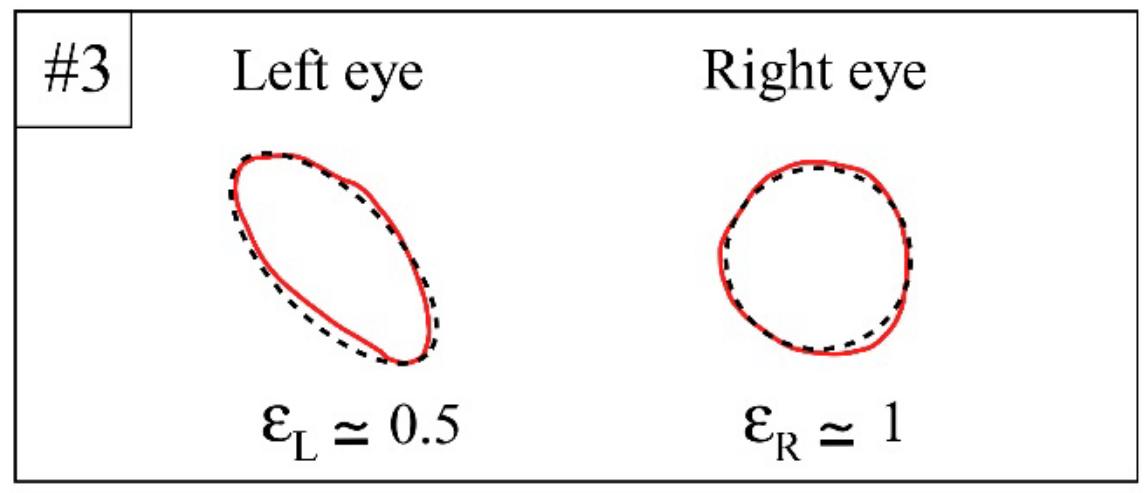

***Figure S1****: Using the foveascope, we have recorded the Maxwell centroid outlines for the three observers. For the observer #1, the quasi-circular left outline determines his left eye dominance and for the observers #2, #3 their quasi-circular outlines for their right eye determine their right eye dominance. The diameters of the centroids for the observers #1, #2, #3 correspond to about 27' ± 3' , 25' ± 3' and 30' ± 3' respectively.*

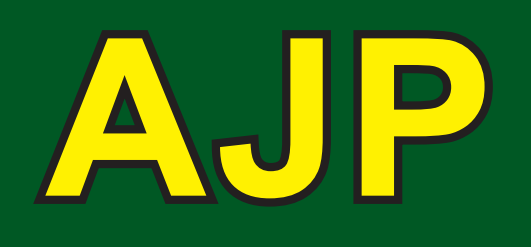

# ASIAN JOURNAL OF PHYSICS

**An International Peer Reviewed Research Journal**

*Special issue in honour of Prof Jay M Enoch*

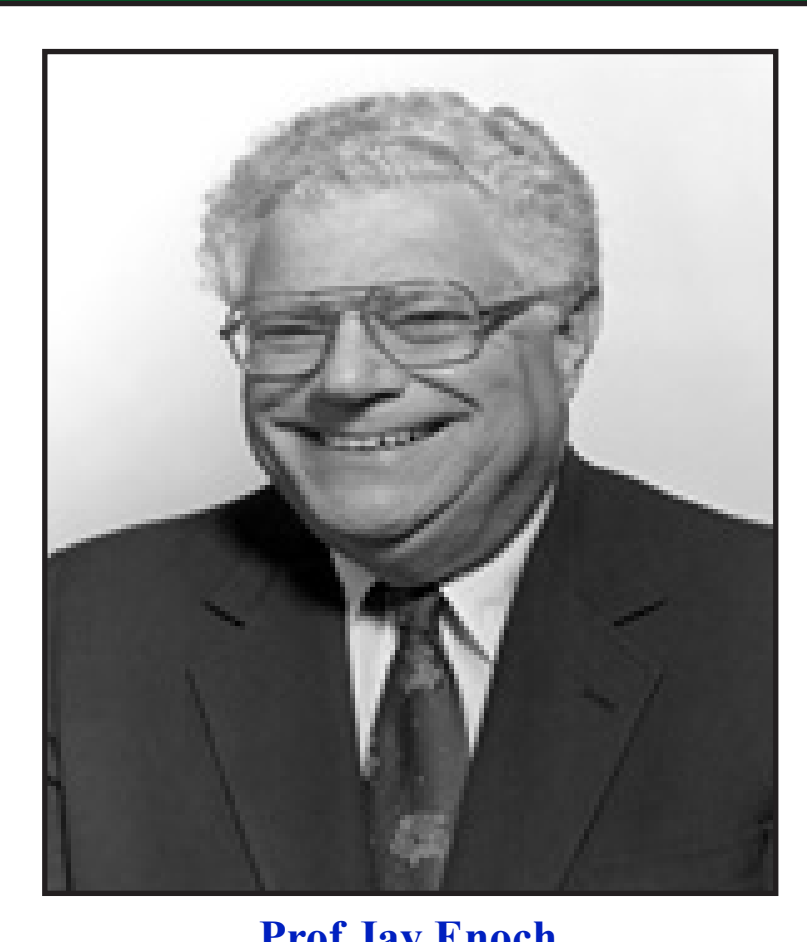

Prof Jay Enoch

*Guest Editors:* **Maria L Calvo & V Lakshminarayanan**

*Co-Guest Editors:* **Shrikant Bharadwaj & Prem Nandhini Satgunam**

# Asian Journal of Physics

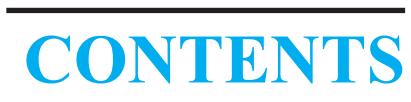

## About Professor Jay Enoch

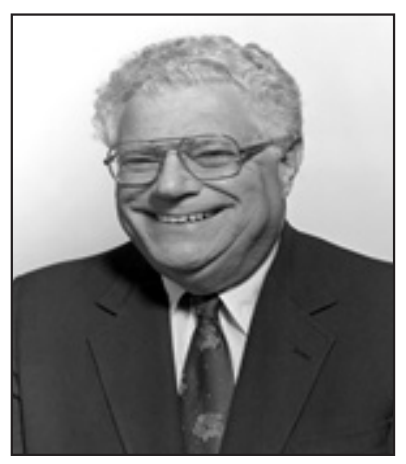

Jay M Enoch

Jay M Enoch was born in April 20, 1929, in New York City. After graduating from Bronx High School of Science in 1946, he completed his pre-optometry studies at Columbia College in 1948 and received his Bachelor Degree in Optometry at Columbia University in 1950. He attended the University of Rochester Institute of Optics in 1953 and obtained his Ph D in Physiological Optics from Ohio State University in 1956. Dr. Enoch served as Post-Doctoral Fellow at National Physical Lab, UK with W S Stiles 1959–60.

Prof Enoch has practiced Optometry in New York, New Jersey, Missouri and Florida, as well as in the US Army. He has received licenses in California, Florida, New York, Missouri, New Jersey and Ohio. His teaching appointments in universities have included: the Ohio State University School of Optometry, the Washington University School of Medicine, the University of Florida College of Medicine, the Waseda University Department of Applied Physics (Tokyo), the University of Bologna School of Medicine and the University of Modena School of Medicine (Italy), the Optics Department in Complutense University of Madrid (Spain) and the University of California School of Medicine, San Francisco.

Jay M Enoch is currently retired as Professor of Optometry and Vision Science. He was Dean Emeritus, at the School of Optometry, University of California at Berkeley (term 1980-1992). For more than sixty years he has achieved important research works and he is considered a pioneer in the field of vision science. Enoch's publications include eight books, over twenty chapters, and more than 300 papers or reports in his areas of expertise and research: perimetry, contact lenses, aniseikonia, psychophysics, hyperacuity testing in dense cataracts, the Stiles-Crawford effect and studies of retinal photoreceptor optics. In addition, he has dedicated time and efforts to contribute on very key aspects of the history of science and optics: in particular, to the history of lenses and mirrors, in a vast period of antique civilizations.

More specifically, his research interests can be summarized as follows:

a) Effects on vision resulting from anomalies of the photo-receptor orientation. He demonstrated that there were effects on vision resulting from anomalies of photo-receptor orientation through measurements of the Stiles-Crawford Effect (directional sensitivity of the retina) in a number of disorders and diseases. As a very important contribution, he demonstrated first that vertebrate photo-receptors are waveguides and characterized their properties in several species. He has studied a trans-retinal tractional effect, as well as the sources of these meaningful traction effects both experimentally and through the use of special models.

b) Regarding optimizing vision in eyes of patients with "*front of the eye*" low vision problems, he developed techniques to correct the vision in neonates and premature infants born with a variety of anterior eye congenital anomalies.

c) In the area of perimetric methods for testing visual fields, he investigated the effects of image blur upon perimetric measurements, developed a perimetry of neuropsychiatric disorders (apparently affected by anomalies of neurotransmitter substances/therapeutic interventions affecting the neurotransmitters), led advances in perimetric and visual field standards, and defined the so-called "layer-by-layer quantitative perimetric techniques."

d) He applied measurements of hyperacuity (Vernier acuity, doubling measurements, bisection tasks, Vernier fields) to clinical practice. He has also been one of a small number of scientists working in this area who have demonstrated that certain Vernier acuity test designs are not affected by age.

Among his most outstanding international activities, Prof Enoch led the development of a quality School of Optometry in Chennai (Madras), India. The Elite School of Optometry is offering, since 1985, a graduate program in Optometry. An external program of the Birla Institute for Technology and Science (Pilani, Rajasthan) has been developed as well. This School is a model for ca. 6 new Schools of Optometry in India (including Mumbai, Puna, Calcutta, Hyderabad). And he has maintained a research laboratory at the Aravind Eye Hospital at Madurai, India.

Dr Enoch's honors and awards include the first (modern series) Glenn Fry Lecture Award (1971), the Charles F Prentice Medal (1974), the Francis I Proctor Medal of ARVO (1977), a Honor Award from the American Academy of Ophthalmology (1985), the Otto Wichterle Medal (1986), Honorary Doctor of Science Degree from the State University of New York (1993), and in 2002 a Honorary Degree from the Polytechnical University of Catalunya (Spain), as the first recipient of such recognition in Optics and Vision. He has been appointed Alameda and Contra Costa Counties Optometric Association Optometrist of the Year in 1988, and was given the Berkeley Citation at his Festshrift in 1996. He was the first optometrist to receive the Pisart Award of the Lighthouse International (2001).

His professional memberships include Association for Research in Vision and Ophthalmology (ARVO), for which he served as Trustee and President, Optical Society of America (Fellow), American Academy of Optometry, American Association for the Advancement of Science (Fellow), and Honorary Life Member, University of California Optometry Alumni Association.

## Editorial

We have immense pleasure in presenting to the readers this special issue of the AJP dedicated on Prof Jay M Enoch who has devoted time and efforts to contribute on very key aspects of the history of science and optics: in particular, to the history of lenses and mirrors, in a vast period of antique civilizations.

Sometime in late Oct 2022, Editor-in-Chief Professor Vinod Rastogi invited us to guest edit an issue of the Asian Journal of Physics in 2023, which we readily accepted whole-heartedly and suggested him that the proposed issue will be dedicated to Prof Jay M Enoch for his significant contributions to the field of Vision Science, which encompasses many disciplines, including optometry, ophthalmology, molecular genetics, neuroscience, and physiological optics.

The present issue of Asian J Phys contains 12 articles covering history of optics, optometry and visión science. We thank all the authors who accepted our invitation and submitted their manuscripts in time. Without their cooperation the present issue would not have seen the light of the day. We may mention that all the papers included in this issue, though were invited, have gone through peer review before getting accepted.

Finally, we would like to thank the editorial team for their tremendous efforts during this period.

We are also grateful to Er Manoj, Er (Ms) Nidhi, Ms Ashmita Singh and Ms Laxmi for all the hardwork they have done toward a successful edition of the special issue dedicated to Jay Enoch.

The issue has papers of interest, and we are extremely pleased to submit it for general readership.

Maria L Calvo
V Lakshminarayan
Shrikant Bharadwaj
PremNandhini Satgunam